\documentclass[aps,prb,a4paper,10pt,twocolumn,amsmath,showpacs,longbibliography,superscriptaddress,bibnotes]{revtex4-2} 
\usepackage{bm}
\usepackage{float}
\usepackage{dcolumn,graphicx,color,booktabs,microtype,afterpage}
\usepackage{amssymb}
\usepackage{amsmath}
\usepackage[charter,greekuppercase=italicized]{mathdesign}
\usepackage{sidecap}
\usepackage[mathlines]{lineno}
\graphicspath{{./}{figure/}}
\DeclareGraphicsExtensions{.eps,.png,.pdf,.jpg}

\renewcommand{\tablename}{Table}
\makeatletter\renewcommand{\fnum@figure}[1]{\figurename~\thefigure.~}\makeatother
\makeatletter\renewcommand{\fnum@table}[1]{\tablename~\thetable.}\makeatother

\newcount\hh \newcount\mm
\hh=\time \divide\hh by 60
\mm=\hh \multiply\mm by 60 \mm=-\mm
\advance\mm by \time
\def\now{\number\hh:\ifnum\mm<10{}0\fi\number\mm}

\usepackage[colorlinks,plainpages=false,linkcolor=blue,urlcolor=blue,citecolor=blue,pdfpagemode=UseNone,pdfstartview=FitBH]{hyperref}

\newcommand{\tcr}{\textcolor{black}}

\SetExpansion[context = sloppy,shrink = 60]{encoding = {OT1,T1,TS1} }{} 

\begin{document}

\makeatletter\renewcommand{\ps@plain}{%
\def\@evenhead{\hfill\itshape\rightmark}%
\def\@oddhead{\itshape\leftmark\hfill}%
\renewcommand{\@evenfoot}{\hfill\small{--~\thepage~--}\hfill}%
\renewcommand{\@oddfoot}{\hfill\small{--~\thepage~--}\hfill}%
}\makeatother\pagestyle{plain}


%
\title{\tcr{Spin order and fluctuations in the EuAl$_4$ and EuGa$_4$ topological antiferromagnets: A $\mu$SR study}}

\author{X. Y.\ Zhu}\thanks{These authors contributed equally}
\affiliation{Key Laboratory of Polar Materials and Devices (MOE), School of Physics and Electronic Science, East China Normal University, Shanghai 200241, China}
\author{H. Zhang}\thanks{These authors contributed equally}
\affiliation{Key Laboratory of Polar Materials and Devices (MOE), School of Physics and Electronic Science, East China Normal University, Shanghai 200241, China}
\author{D.\ J.\ Gawryluk}
\affiliation{Laboratory for Multiscale Materials Experiments, Paul Scherrer Institut, Villigen CH-5232, Switzerland}
\author{Z.\ X.\ Zhen}
\affiliation{Key Laboratory of Polar Materials and Devices (MOE), School of Physics and Electronic Science, East China Normal University, Shanghai 200241, China}
\author{B.\ C.\ Yu}
\affiliation{Key Laboratory of Polar Materials and Devices (MOE), School of Physics and Electronic Science, East China Normal University, Shanghai 200241, China}
\author{S.\ L.\ Ju}
\affiliation{Swiss Light Source, Paul Scherrer Institut, Villigen CH-5232, Switzerland}

\author{W.\ Xie}
\affiliation{DESY, Notkestra$\beta$e 85, D-22607 Hamburg, Germany}
\author{D.\ M.\ Jiang}
\affiliation{Key Laboratory of Polar Materials and Devices (MOE), School of Physics and Electronic Science, East China Normal University, Shanghai 200241, China}
\author{W.\ J.\ Cheng}
\affiliation{Key Laboratory of Polar Materials and Devices (MOE), School of Physics and Electronic Science, East China Normal University, Shanghai 200241, China}
\author{Y.\ Xu}
\affiliation{Key Laboratory of Polar Materials and Devices (MOE), School of Physics and Electronic Science, East China Normal University, Shanghai 200241, China}
\author{M.\ Shi}
\affiliation{Swiss Light Source, Paul Scherrer Institut, Villigen CH-5232, Switzerland}
\author{E.\ Pomjakushina}
\affiliation{Laboratory for Multiscale Materials Experiments, Paul Scherrer Institut, Villigen CH-5232, Switzerland}

\author{Q.\ F.\ Zhan}
\affiliation{Key Laboratory of Polar Materials and Devices (MOE), School of Physics and Electronic Science, East China Normal University, Shanghai 200241, China}
\author{T.\ Shiroka}\email[Corresponding authors:\\]{tshiroka@phys.ethz.ch}
\affiliation{Laboratory for Muon-Spin Spectroscopy, Paul Scherrer Institut, Villigen PSI, Switzerland}
\affiliation{Laboratorium f\"ur Festk\"orperphysik, ETH Z\"urich, CH-8093 Z\"urich, Switzerland}
\author{T.\ Shang}\email[Corresponding authors:\\]{tshang@phy.ecnu.edu.cn}
\affiliation{Key Laboratory of Polar Materials and Devices (MOE), School of Physics and Electronic Science, East China Normal University, Shanghai 200241, China}

\date{\today}

\begin{abstract}
EuAl$_4$ and EuGa$_4$ are two candidate materials for studying 
the interplay between 
correlated-electron phenomena, topological spin textures, and topologically nontrivial bands.
Both compounds crystallize in a centrosymmetric tetragonal BaAl$_4$-type structure 
(space group $I4/mmm$) and show antiferromagnetic (AFM) order 
below $T_\mathrm{N} = 15.6$ and 16.4\,K, respectively.
Here, we report on systematic muon-spin rotation and relaxation ($\mu$SR) 
studies of the magnetic properties of EuAl$_4$ and EuGa$_4$ single crystals 
at a microscopic level. 
Trans\-verse\--field $\mu$SR measurements, spanning a wide temperature 
range (from 1.5 to 50\,K), show clear bulk AFM transitions, with an 
almost 100\%  magnetic volume fraction in both cases.
Zero-field $\mu$SR measurements, covering both the AFM and the paramagnetic 
(PM) states, reveal 
internal magnetic fields $B_\mathrm{int}(0) = 0.33$\,T and 
0.89\,T in EuAl$_4$ and EuGa$_4$, respectively.
The transverse muon-spin relaxation rate $\lambda_\mathrm{T}$, a measure 
of the internal field distribution at the muon-stopping site, 
shows a contrasting behavior. In EuGa$_4$, it decreases with lowering 
the temperature, reaching its minimum at zero temperature, 
$\lambda_\mathrm{T}(0) = 0.71$\,$\mu$s$^{-1}$. 
In EuAl$_4$, it increases 
significantly below $T_\mathrm{N}$, to reach 58\,$\mu$s$^{-1}$ at \tcr{1.5\,K},
most likely reflecting the complex magnetic 
structure and the competing interactions in the AFM state of EuAl$_4$.   
In both compounds, the temperature-dependent longitudinal muon-spin relaxation 
$\lambda_\mathrm{L}(T)$, an indication of the rate of 
spin fluctuations, diverges near the onset of AFM order, followed by 
a significant drop at $T < T_\mathrm{N}$. 
In the AFM state, spin fluctuations are much stronger in 
EuAl$_4$ than in EuGa$_4$, while being comparable in the PM state. 
The evidence of \tcr{robust} spin fluctuations against \tcr{the external 
magnetic fields} provided by $\mu$SR
may offer new 
insights into  the origin of the topological Hall effect and the 
possible magnetic skyrmions in the EuAl$_4$ and EuGa$_4$ compounds.
\end{abstract}

\maketitle

\section{\label{sec:Intro}Introduction}\enlargethispage{8pt}
Topological materials are at the forefront of quantum matter and material science research due to their great potential for applications~\cite{Armitage2018,Lv2021}.
Recently, the discovery of nontrivial band topology and extremely large magnetoresistance in the BaAl$_4$ compound has stimulated considerable interest in this family of materials~\cite{wang_crystalline_2021}.
The tetragonal BaAl$_4$-type structure with a space group of $I4/mmm$ (No.~139) represents the prototype for many binary- and ternary derivative compounds~\cite{Kneidinger2014}, as e.g., heavy-fermion compounds and iron-based high-$T_c$ superconductors.
 
Upon replacing Ba with Sr or Eu, or when replacing Al with Ga, 
all AE(Al,Ga)$_4$ (AE = Sr, Ba, and Eu) crystallize in the same tetragonal structure, while Ca(Al,Ga)$_4$ adopts a 
monoclinic crystal structure with a space group $C2/m$ (No.~12)~\cite{wang_crystalline_2021,Nakamura2016,Nakamura2016}. 
\tcr{Among these materials, the Eu-4$f$ electrons bring new intriguing aspects to the topology}.
Both EuAl$_4$ and EuGa$_4$ are antiferromagnets below their critical temperatures $T_\mathrm{N}$ = 15.6, and 16.4\,K, respectively, with the former also undergoing a CDW transition at $T_\mathrm{CDW} \sim 140$\,K~\cite{EuAl4_PRB,Shang2021,araki_charge_2013,nakamura_unique_2014,nakamura_transport_2015,shimomura_lattice_2019,Kobata2016}. 
Further, while EuGa$_4$ exhibits only one antiferromagnetic (AFM) transition, EuAl$_4$ undergoes four subsequent AFM transitions below $T_\mathrm{N}$. 
More interestingly, by applying a magnetic field along the $c$-axis, 
both EuAl$_4$ and EuGa$_4$ undergo a series of metamagnetic transitions in the AFM state~\cite{EuAl4_PRB,Shang2021,nakamura_transport_2015}.  
Within a field range of $\sim$1--2.5\,T (EuAl$_4$) or $\sim$4--7\,T (EuGa$_4$), 
a clear hump-like anomaly is observed in the Hall resistivity, 
most likely a manifestation of the topological Hall effect (THE)~\cite{EuAl4_PRB,Shang2021}. 
Very recently, a THE has been observed also in Al-doped EuGa$_4$~\cite{Moya2021}, 
which exhibits comparable critical fields to EuAl$_4$~\cite{EuAl4_PRB}. 

The topological Hall effect is considered to be the hallmark of spin textures with a finite scalar spin chirality.
Such topological spin textures usually exhibit a nonzero Berry phase, here acting as an effective magnetic field, giving rise
to the topological Hall resistivity~\cite{Tokura2021}. THE is frequently
observed in magnetic materials with non-coplanar spin textures, such as magnetic skyrmions~\cite{neubauer_topological_2009,gayles_dzyaloshinskii-moriya_2015,kanazawa_large_2011,franz_real-space_2014,kurumaji_skyrmion_2019,lee_unusual_2009,li_robust_2013,huang_extended_2012,schulz_emergent_2012,qin_emergence_2019,matsuno_interface-driven_2016}.
Skyrmions are one of the most intriguing topologically nontrivial spin textures that can be easily manipulated~\cite{Jonietz2010}, hence holding a promise
for diverse applications, such as high-density spintronics~\cite{nagaosa_topological_2013,Fert2017}.
\tcr{THE has been observed mostly in magnetic compounds whose 
crystal structure lacks an inversion center, while 
centrosymmetric compounds that host magnetic skyrmions are 
rare~\cite{neubauer_topological_2009,gayles_dzyaloshinskii-moriya_2015,kanazawa_large_2011,franz_real-space_2014,kurumaji_skyrmion_2019,Hirschberger2019,Khanh2020}.}
Eu(Al,Ga)$_4$ represent such rare cases where to look for 
the possible existence of magnetic skyrmions~\cite{EuAl4_PRB,Shang2021,Moya2021}.   
\tcr{According to neutron diffraction studies}, 
in the AFM state, the magnetic $q$-vector of EuAl$_4$ changes from $\bm{q}_1$ = (0.085, 0.085, 0) at $T_\mathrm{N}$ = 13.5\,K to $\bm{q}_2$ = (0.170, 0, 0) at 11.5\,K and slightly to $\bm{q}_3$ = (0.194, 0, 0) at 4.3\,K~\cite{Kaneko2021}. 
Unlike the complex incommensurate
transitions observed 
in EuAl$_4$, the AFM structure of EuGa$_4$ is
described by a simple $\bm{q}$ = (0, 0, 0) magnetic vector,
with the Eu moments lying in the basal $ab$-plane~\cite{Kawasaki2016}. 
Noncollinear spins with incommensurate propagation vectors have been reported also in the isostructural EuGa$_2$Al$_2$~\cite{Moya2021}.

\begin{figure}[!thb]
	\centering
	\includegraphics[width=0.49\textwidth]{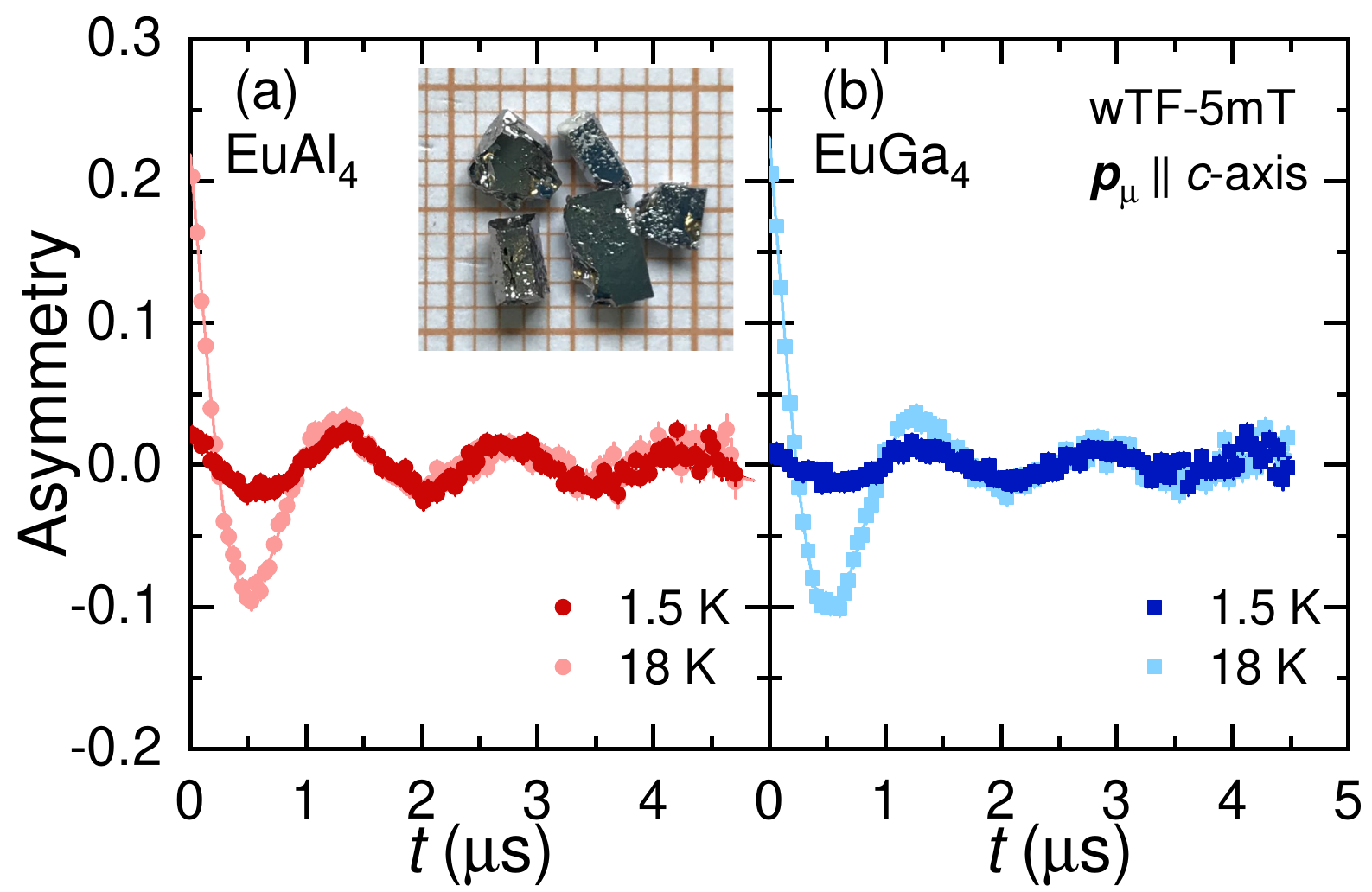} 
	\caption{\label{fig:wTF}Time-domain wTF-$\mu$SR spectra of (a) EuAl$_4$ 
	and (b) EuGa$_4$ single crystals, collected in the AFM (1.5\,K) and PM (18\,K) states in a weak transverse field of 5\,mT. 
	The solid lines represent fits to Eq.~\eqref{eq:wTF}. The inset in (a) 
	depicts the EuAl$_4$ crystals, aligned with their $c$ axis 
	parallel to the muon momentum direction, i.e., 
	$\bm{p}_{\mu}$ $\parallel$ $c$.}
\end{figure}

As an extremely sensitive magnetic probe at a microscopic level, 
the muon-spin rotation and relaxation ($\mu$SR) technique lends itself naturally to studying
the temperature evolution of the magnetic properties of EuAl$_4$ and 
EuGa$_4$ single crystals. 
As shown in detail below, we report: i) the intrinsic
fields at the muon implantation sites in EuAl$_4$ and EuGa$_4$ across 
the respective phase diagrams in the absence of external magnetic fields; 
ii) the magnetic volume fraction in the AFM state; iii) evidence of strong spin fluctuations. 

\section{Experimental details\label{sec:details}}\enlargethispage{8pt}

Single crystals of EuAl$_4$ and EuGa$_4$ were grown by a molten Al- and 
Ga flux method, respectively, the details of growth being reported 
elsewhere~\cite{EuAl4_PRB,Shang2021}. 
The crystal orientation was checked by x-ray diffraction (XRD) measurements using a Bruker D8 diffractometer with Cu K$_\alpha$ radiation.
The magnetic susceptibility measurements were performed on a Quantum Design magnetic properties measurement system (MPMS) with the applied magnetic field along the $c$-axis. 

$\mu$SR experiments were carried out at the general-purpose surface-muon
(GPS) instrument at the $\pi$M3 beam line of the Swiss muon source (S$\mu$S) at Paul Scherrer Institut (PSI) in Villigen, Switzerland. 
In this study, we performed three kinds of experiments: weak transverse-field (wTF)-$\mu$SR, zero-field (ZF)-, and longitudinal-field (LF)-$\mu$SR measurements.
As to the former, we could determine the temperature evolution of the 
magnetic volume fraction. As to the latter two, we aimed at studying the 
temperature evolution of the magnetically ordered phase and the dynamics of spin fluctuations.  

The aligned EuAl$_4$ and EuGa$_4$ crystals were positioned on a thin 
aluminum tape, with their $c$-axes parallel to the muon\--mo\-mentum direction, 
i.e., $\bm{p}_{\mu}$ $\parallel$ $c$ [see inset in Fig.~\ref{fig:wTF}(a)].
For the wTF-$\mu$SR measurements, the applied magnetic field $B_\mathrm{appl}$ 
was perpendicular to the muon-spin direction 
(i.e., $B_\mathrm{appl} \perp \bm{S}_{\mu}$), while it was 
parallel for the LF-$\mu$SR measurements 
(i.e., $B_\mathrm{appl} \parallel \bm{S}_{\mu}$). In both wTF- and LF-$\mu$SR cases, the crystals were cooled in an applied 
magnetic field down to the base temperature (i.e., 1.5 K). 
For the ZF-$\mu$SR measurements, to exclude the possibility of 
stray magnetic fields, the magnets were degaussed before the 
measurements. All the $\mu$SR spectra were collected upon heating and were analyzed by means of the \texttt{musrfit} software package~\cite{Suter2012}.

\begin{figure}[!tbh]
	\centering
	\includegraphics[width=0.48\textwidth]{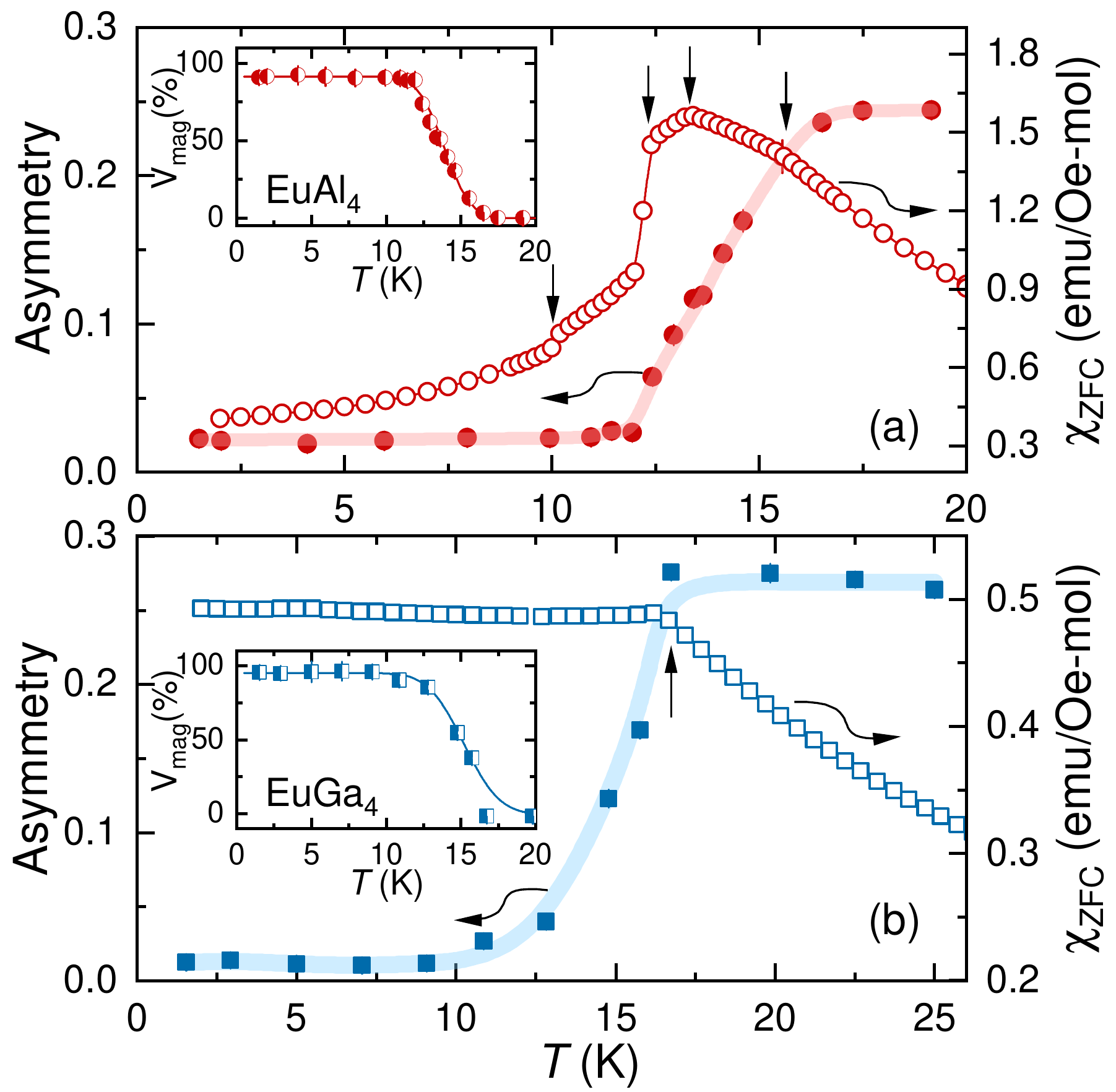}
	\caption{\label{fig:Vmag}Temperature dependence of the $A_\mathrm{NM}$ 
	asymmetry (left-axis) of wTF-$\mu$SR spectra for (a) EuAl$_4$ and 
	(b) EuGa$_4$ single crystals. We report also the 
	magnetic susceptibilities $\chi_\mathrm{ZFC}$ (right axes), measured 
	in a field of $\mu_0H = 0.1$\,T after zero-field cooling (ZFC). 
	In both cases, the insets show the magnetic volume fraction vs 
	temperature. Here, lines are fits to a phenomenological function 
	[see Eq.~\eqref{eq:erf}]. The vertical arrows mark the AFM transitions. 
	The magnetic susceptibility data were taken from Refs.~\onlinecite{EuAl4_PRB,Shang2021}.}
\end{figure}

\section{Results and discussion\label{sec:results}}\enlargethispage{8pt}
\subsection{wTF-$\mu$SR}%
The magnetic transition temperatures $T_\mathrm{N}$ and the
evolution with temperature of the magnetic volume fraction
in EuAl$_4$ and EuGa$_4$ single crystals were
established by means of wTF-$\mu$SR measurements. A weak transverse field of 5\,mT was applied perpendicular to the initial muon-spin direction 
in the PM state, 
where it leads to oscillations, as shown in Fig.~\ref{fig:wTF}. 
In the long-range ordered AFM state (i.e., 1.5\,K), the applied 
5-mT field is much smaller than the internal fields. As a consequence, 
upon entering the AFM state, muon spins precess with frequencies 
that reflect the internal fields at the muon-stopping sites rather than 
the weak applied field. Normally, the magnetic order leads to a very 
fast muon-spin depolarization in the first tenths of $\mu$s 
(see also the ZF-$\mu$SR spectra in the insets of Fig.~\ref{fig:ZF_muSR}).
Therefore, the wTF-$\mu$SR spectra can be described by the function:   
\begin{equation}
	\label{eq:wTF}
	A_\mathrm{wTF}(t) = A_\mathrm{NM} \cos(\gamma_{\mu} B_\mathrm{int} t + \phi) \cdot e^{-\lambda t}, 	
\end{equation}
where $A_\mathrm{NM}$ is the initial muon-spin asymmetry (i.e., the 
amplitude of the oscillation) for muons implanted in the nonmagnetic 
(NM) or PM fraction of EuAl$_4$ and EuGa$_4$ single crystals;  
$\gamma_{\mu}$$B_\mathrm{int}$ is the muon-spin precession frequency,  
with $\gamma_{\mu} = 2\pi \times 135.5$\,MHz/T the muon gyromagnetic ratio  
and $B_\mathrm{int}$ the local field sensed by muons (here
almost identical to the applied magnetic field, i.e., $B_\mathrm{int} \sim 5$\,mT); 
$\phi$ is the initial phase, and $\lambda$ is the muon-spin relaxation rate.
Note that, in the AFM state, the very fast $\mu$SR relaxation was excluded   
and only the residual slow-relaxing asymmetry was analyzed 
(see the 1.5-K dataset in Fig.~\ref{fig:wTF}).

Figure~\ref{fig:Vmag} summarizes the resulting wTF-$\mu$SR asymmetry values $A_\mathrm{NM}$ as a function 
of temperature. In the PM state, all the implanted muons precess at the same frequency $\gamma_{\mu}$$B_\mathrm{int}$. As the temperature approaches $T_\mathrm{N}$, only the muons implanted 
in the remaining PM/NM phase precess at the frequency $\gamma_{\mu}$$B_\mathrm{int}$, 
here reflected in a reduced oscillation amplitude.  
The PM (or NM) sample fraction is determined from the oscillation 
amplitude. In both EuAl$_4$ and EuGa$_4$, 
$A_\mathrm{NM}$ starts to decrease near the onset of AFM order, where 
also the magnetic susceptibilities show clear transitions. Although 
EuAl$_4$ undergoes four successive AFM transitions [indicated by 
vertical arrows in Fig.~\ref{fig:Vmag}(a)], $A_\mathrm{NM}(T)$ does 
not capture them individually,
as it is sensitive only to the global PM (or NM) volume fraction. 
The temperature evolution of the magnetic volume fraction can be derived 
from $V_\mathrm{mag}(T) = 1 - A_\mathrm{NM}(T)/A_\mathrm{NM}(T > T_\mathrm{N})$. 
The $V_\mathrm{mag}(T)$ values are summarized in the insets of Fig.~\ref{fig:Vmag}(a) 
and \ref{fig:Vmag}(b) for EuAl$_4$ and EuGa$_4$, respectively. 
To determine 
the magnetic volume fraction $V_\mathrm{mag}$, the average magnetic 
transition temperature $T_\mathrm{N}$, and the transition width $\Delta T$, 
$V_\mathrm{mag}(T)$ data were fitted using the phenomenological function:               
\begin{equation}
	\label{eq:erf}
	V_\mathrm{mag}(T) = V_\mathrm{mag}(0)\;\frac{1}{2}\left[1 - \mathrm{erf}\left(\frac{T-T_\mathrm{N}}{\sqrt{2}\Delta T}\right)\right],
\end{equation}
where erf($T$) is the error function. As shown by solid lines in 
the insets of Fig.~\ref{fig:Vmag}, for EuAl$_4$, we obtain $T_\mathrm{N} = 13.9(2)$\,K, 
$\Delta T = 1.4(2)$\,K, and $V_\mathrm{mag}(0) = 91(2)\%$; while for 
EuGa$_4$, $T_\mathrm{N} = 15.2(3)$\,K, $\Delta T = 1.8(2)$\,K, and 
$V_\mathrm{mag}(0) = 95(2)\%$. Both samples show sharp transitions and 
can be considered as fully magnetically ordered at low temperatures, 
indicative a 
high sample quality. Note also that, the transition temperatures, as 
determined from $V_\mathrm{mag}(T)$, have their onset at $\sim 16.5$\,K  
and $\sim 16.7$\,K for EuAl$_4$ and EuGa$_4$, both in very good 
agreement with the magnetometry data. 

\subsection{ZF- and LF-$\mu$SR}%

\begin{figure}[tbh]
	\centering
	\vspace{-2mm}
	\includegraphics[width=0.46\textwidth]{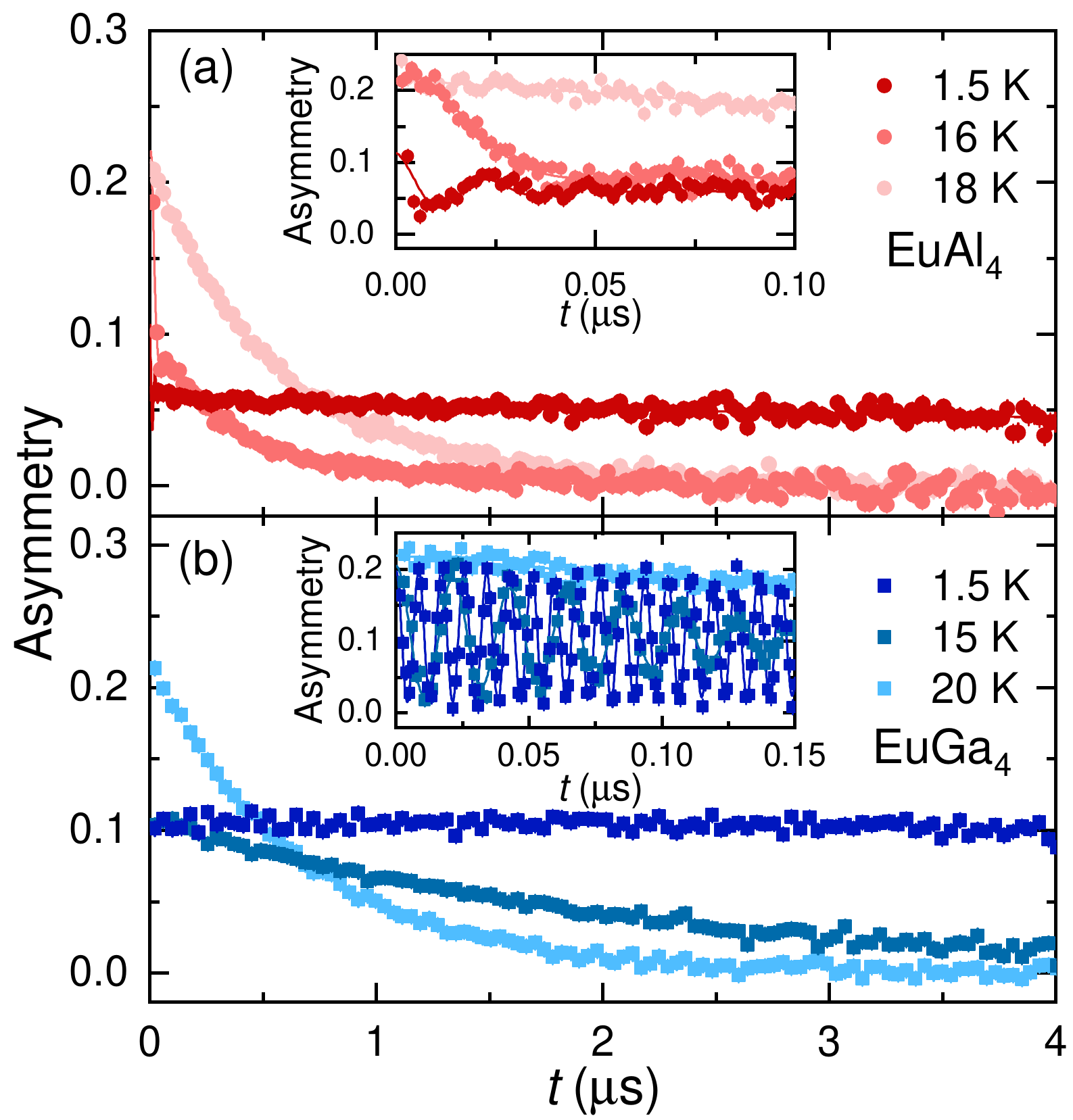}
	\caption{\label{fig:ZF_muSR}Representative ZF-$\mu$SR spectra 
	collected in a transverse muon-spin configuration 
	($\bm{p}_{\mu} \perp \bm{S}_{\mu}$) at temperatures covering both 
	the PM and AFM states for (a) EuAl$_4$ and (b) EuGa$_4$, respectively.
	Insets highlight the short-time spectra, illustrating the 
	coherent oscillations caused by the long-range AFM order. 
	Solid lines through the data are fits to Eq.~\eqref{eq:ZF} (see text for details).
	}
\end{figure}

To investigate the local magnetic order of EuAl$_4$ and EuGa$_4$ single crystals,
ZF-$\mu$SR spectra were collected at different 
temperatures, covering both the PM and AFM states. The time evolution of 
ZF-$\mu$SR asymmetry, $A_\mathrm{ZF}(t)$, encodes the local magnetic fields and their distribution at the muon-stopping site. If the electronic magnetic moments 
fluctuate very fast (typically above $10^{12}$\,Hz in the PM state), 
they do not influence the muon-spin polarization. 
Randomly oriented slow fluctuating or static moments (below $10^4$ Hz, 
such as nuclear spins, or electronic moments in spin glasses), give rise 
to incoherent precessions and a slow depolarization. 
Conversely, in case of ordered static moments, a fast depolarization 
and superimposed oscillations, reflecting the coherent precession of 
the muon spins, are observed~\cite{Yaouanc2011}.
This is clearly demonstrated in Fig.~\ref{fig:ZF_muSR}, where the time 
evolution of selected ZF-$\mu$SR spectra for EuAl$_4$ and EuGa$_4$ are presented.

%

In the PM state ($T > T_\mathrm{N}$), the $\mu$SR spectra still exhibit 
a relatively fast muon-spin depolarization ($\sim 2$\,$\mu$s$^{-1}$), 
implying the existence of strong spin fluctuations, 
here further confirmed by LF-$\mu$SR measurements (see below). 
In absence of spin fluctuations, the muon-spin depolarization is 
usually due to the nuclear dipole fields~\cite{Yaouanc2011}, with a typical value of 
less than 0.1\,$\mu$s$^{-1}$ in EuAl$_4$ and EuGa$_4$~\cite{Fujita2020}.
The $\mu$SR spectra in the AFM state ($T \le T_\mathrm{N}$) are
characterized by highly damped oscillations, typical of long-range magnetic order 
(see insets in Fig.~\ref{fig:ZF_muSR}), superimposed on a slowly decaying 
relaxation, observable only at long times. To track
these changes across the whole
temperature range,
the ZF-$\mu$SR spectra of EuAl$_4$ and EuGa$_4$
were analyzed using the following model: 
\begin{equation}\label{eq:ZF}
	\begin{split}
		A_\mathrm{ZF}(t)  =  &\,  A_1 \cdot \left[\alpha \cos(\gamma_{\mu} B_\mathrm{int} t + \phi) \cdot e^{-\lambda_\mathrm{T} t} + (1 - \alpha) \cdot e^{-\lambda_\mathrm{L} t}\right] + \\
		 & A_2  \cdot e^{-\lambda_\mathrm{tail} t}.
	\end{split}
\end{equation}
Here, $\alpha$ and $1 - \alpha$ are the oscillating (i.e., transverse) 
and nonoscillating (i.e., longitudinal) fractions of the $\mu$SR signal, 
respectively, whose initial total asymmetry is equal to $A_1$.
$\lambda_\mathrm{T}$ and $\lambda_\mathrm{L}$ represent the 
transverse and longitudinal relaxation rates, 
while $A_1$ and $A_2$ represent the asymmetries of the two nonequivalent 
muon-stopping sites. 
In EuAl$_4$, muons stopping at the second site do not undergo any precession, but
show only a slow relaxation, here described by $\lambda_\mathrm{tail}$.  
In EuGa$_4$, a single muon-stopping site is sufficient to describe 
the ZF-$\mu$SR spectra. Finally, $B_\mathrm{int}$, $\phi$, and $\gamma_{\mu}$ are the same as in Eq.~\eqref{eq:wTF}.  
Similar expressions have been used to analyze the $\mu$SR data 
in other Eu-based magnetic materials, most notably, in the Eu122 iron 
pnictides~\cite{Tran2018,Guguchia2013}. 

In polycrystalline materials with a long-range magnetic order, one 
expects $\alpha = 2/3$, since statistically one third of the muon spins 
are aligned parallel to the local field direction 
(i.e., $\bm{S}_{\mu} \parallel B_\mathrm{int}$) and, hence, do not precess. 
In EuAl$_4$ and EuGa$_4$ single crystals, we find $\alpha$ to be 0.87 
and 0.46, respectively. Since the ZF-$\mu$SR spectra were collected in 
a rotated muon-spin configuration (i.e., $\bm{S}_{\mu}$  $\perp$ $\bm{p}_{\mu}$), 
and the $c$-axis is parallel to the muon momentum (i.e., $c \parallel \bm{p}_{\mu}$), the internal
%
\begin{figure}[bht]
	\centering
	\includegraphics[width=0.45\textwidth]{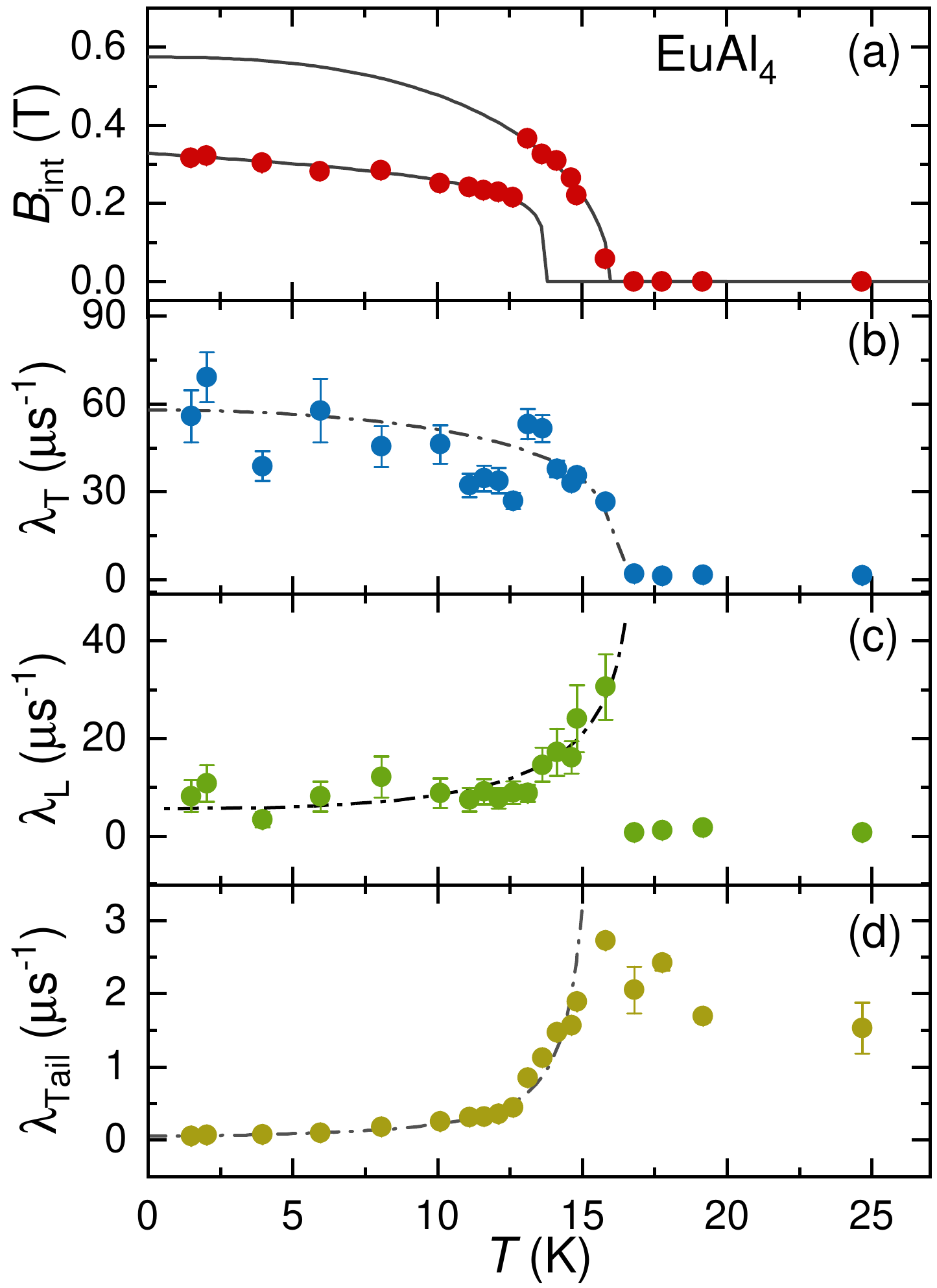}
	\caption{\label{fig:ZF-EA}Temperature dependence of (a) internal field $B_\mathrm{int}(T)$, (b) transverse muon-spin relaxation rate (known also as damping rate) $\lambda_\mathrm{T}(T)$, and (c) longitudinal muon-spin relaxation rate  $\lambda_\mathrm{L}(T)$ for EuAl$_4$, 
	as derived from ZF-$\mu$SR analysis. The muon-spin relaxation rate  
	of the tail of the ZF-$\mu$SR spectra, $\lambda_\mathrm{tail}$,  
	is shown in panel (d).
	Solid lines are fits to the equations described in the text; 
	\tcr{dash-dotted lines are guides to the eyes}.} 
\end{figure}
%
magnetic fields at the muon-stopping sites should be mostly aligned 
along the [001]-direction in EuAl$_4$, but along the [111]-direction in EuGa$_4$. 

\begin{figure}[bht]
	\centering
	\includegraphics[width=0.45\textwidth]{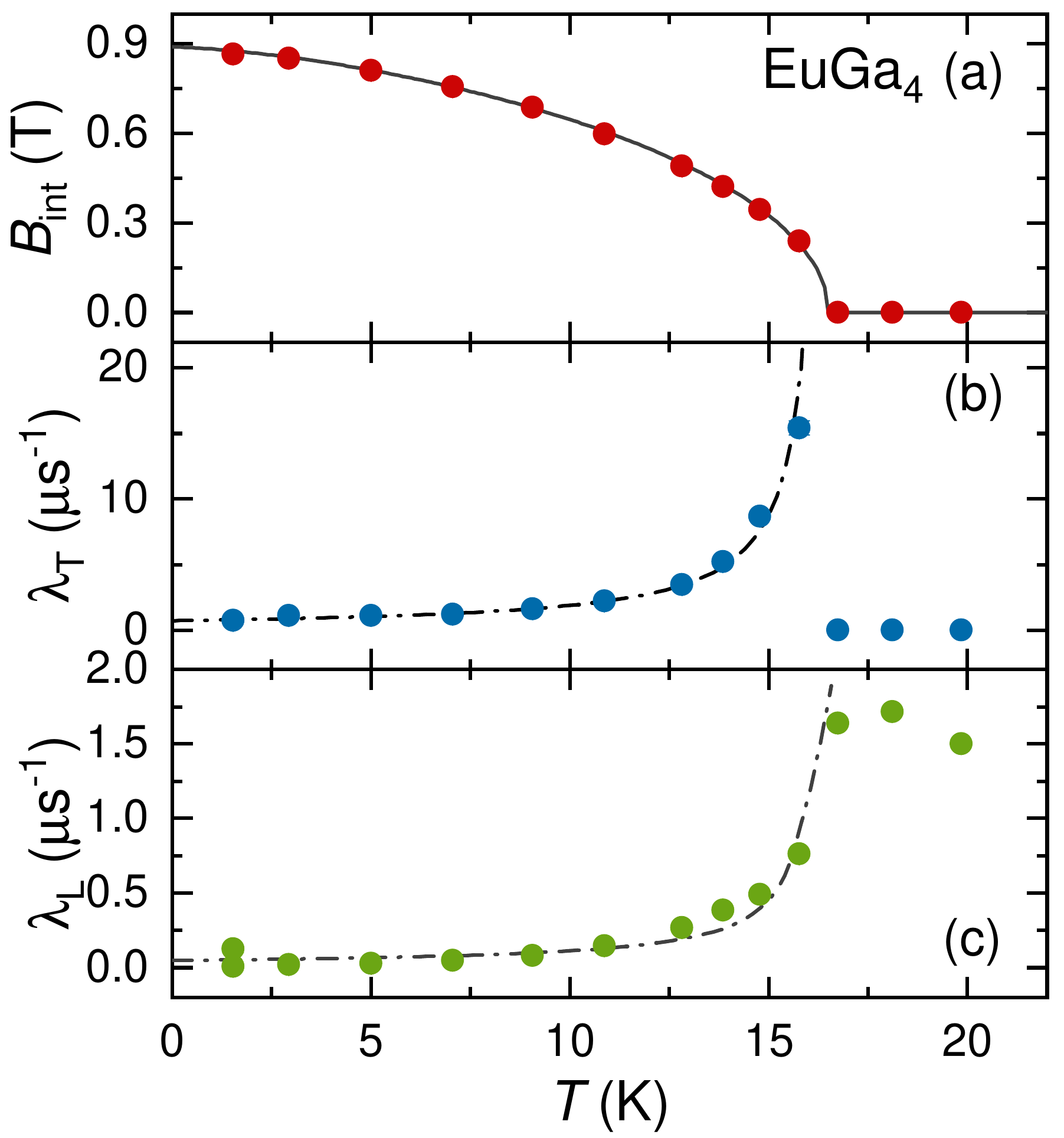}
	\caption{\label{fig:ZF-EG}Temperature dependence of (a) internal field $B_\mathrm{int}(T)$, (b) transverse muon-spin relaxation rate (i.e., damping rate) $\lambda_\mathrm{T}(T)$, 
		and (c) longitudinal muon-spin relaxation rate $\lambda_\mathrm{L}(T)$ 
		for EuGa$_4$, as derived from ZF-$\mu$SR analysis. 	
		Solid lines are fits to the equations described in the text; 
		\tcr{dash-dotted lines are guides to the eyes}.}
\end{figure}

The derived fit parameters for both cases are summarized in 
Fig.~\ref{fig:ZF-EA} and Fig.~\ref{fig:ZF-EG}. 
As can be clearly seen in the top panels,
EuAl$_4$ and EuGa$_4$ show rather different
$B_\mathrm{int}(T)$ behaviors.   
In EuAl$_4$, the $B_\mathrm{int}(T)$ undergoes a sudden drop at $\sim$13\,K, 
which corresponds to the second AFM transition in the magnetic susceptibility 
[see Fig.~\ref{fig:Vmag}(a)].  
Conversely, in EuGa$_4$, $B_\mathrm{int}(T)$ resembles the typical 
mean-field type curve below $T_\mathrm{N}$. Since $B_\mathrm{int}$ is 
directly proportional to 
the magnetic moment, the evolution of $B_\mathrm{int}$ reflects that of the magnetic structure. According to neutron scattering 
studies, in EuAl$_4$, the magnetic $q$-vector changes from $\bm{q}_1$ = (0.085, 0.085, 0) at $T_\mathrm{N}$ = 13.5\,K to $\bm{q}_2$ = (0.170, 0, 0) at 11.5\,K and slightly to $\bm{q}_3$ = (0.194, 0, 0) at 4.3\,K~\cite{Kaneko2021}. Therefore, we identify the drop of $B_\mathrm{int}$ 
at 13\,K with the critical temperature where the magnetic structure changes 
from  $q_1$ to $q_2$. At the same time, the modification of magnetic 
structure from $q_2$ to $q_3$ is too tiny to have a measurable effect on $B_\mathrm{int}$. 
By contrast, the AFM structure of EuGa$_4$ is rather simple [its 
magnetic vector being $\bm{q}$ = (0, 0, 0)] and it 
persists down to 2\,K~\cite{Kawasaki2016}.
As a consequence, in EuGa$_4$, $B_\mathrm{int}$ decreases monotonically 
as the temperature increases. 
In both compounds, $B_\mathrm{int}(T)$ can be modeled by the 
phenomenological equation:
\begin{equation}\label{eq:Bint}
		B_\mathrm{int}(T)  =  B_\mathrm{int}(0) \cdot \left[1 - \left(\frac{T}{T_\mathrm{N}}\right)^{\gamma} \right]^{\delta}.
\end{equation}
Here, $B_\mathrm{int}(0)$ is the internal magnetic field at zero temperature, 
while $\gamma$ and $\delta$ are two empirical parameters.  
%
%
%
%
As indicated by the solid lines in Fig.~\ref{fig:ZF-EA}(a) and 
Fig.~\ref{fig:ZF-EG}(a), the above model describes the data reasonably well,  
yielding the parameters listed in Table~\ref{tab:parameters}.  
In EuAl$_4$, the first AFM phase (AFM1) is characterized by 
$B_\mathrm{int}(0) = 0.57(5)$\,T. The change in magnetic structure lowers 
$B_\mathrm{int}(0)$ down to 0.33(2)\,T in the second AFM phase (AFM2). 
In EuGa$_4$, $B_\mathrm{int}(T)$ follows the typical mean-field type 
curve, yielding $B_\mathrm{int}(0) = 0.89(2)$\,T. 
Considering the presence of the same magnetic Eu$^{2+}$ ions in both cases 
and the similar lattice parameters,  
\tcr{the significantly different $B_\mathrm{int}(0)$ values are most 
likely attributed
to the different muon-stopping sites or to different magnetic 
structures in EuAl$_4$ and EuGa$_4$, the latter having been proved by 
neutron scattering studies. Indeed, at base temperature, EuAl$_4$ 
exhibits a  
complex incommensurate magnetic structure, while this is 
commensurate in EuGa$_4$~\cite{Kaneko2021,Kawasaki2016}.}

The temperature dependence of the transverse and longitudinal $\mu$SR relaxation rates $\lambda_\mathrm{T}(T)$ and $\lambda_\mathrm{L}(T)$ are summarized in Figs.~\ref{fig:ZF-EA}(b) and (c) for EuAl$_4$ and in 
Figs.~\ref{fig:ZF-EG}(b) and (c) for EuGa$_4$, respectively. The transverse relaxation rate $\lambda_\mathrm{T}$ is a measure of the width of static magnetic field distribution at the muon-stopping site
and is also affected by dynamical effects, as e.g., spin fluctuations. 
The longitudinal relaxation rate $\lambda_\mathrm{L}$ is determined 
solely by spin fluctuations.
In EuAl$_4$ and EuGa$_4$, $\lambda_\mathrm{T}(T)$ exhibits completely 
opposite behaviors. In EuAl$_4$ [see Fig.~\ref{fig:ZF-EA}(b)], $\lambda_\mathrm{T}$ is zero in the PM state, and becomes increasingly prominent as the temperature decreases below $T_\mathrm{N}$, reflecting a more disordered field
distribution well inside the AFM state.
Such a large $\lambda_\mathrm{T}$ at temperatures far below $T_\mathrm{N}$ is unusual for an antiferromagnet, and implies an increasingly inhomogeneous distribution of local fields in the AFM state of EuAl$_4$.
\tcr{Thus, at 1.5\,K, $\lambda_\mathrm{T} \sim 58(10)$\,$\mu$s$^{-1}$, 
which implies a half-width at half-maximum (HWHM) 
of field distribution $\Delta = 68(12)$\,mT (here, 
$\Delta = \lambda_\mathrm{T}/\gamma_{\mu}$).}
Such enhanced \tcr{local-field distribution} might be related to the complex spatial arrangement
of the Eu magnetic moments in EuAl$_4$, where the magnetic propagation
vector is incommensurate with the crystal lattice~\cite{Kaneko2021}.
By contrast, in EuGa$_4$, $\lambda_\mathrm{T}(T)$ follows
the typical behavior of materials with a long-range (anti)ferromagnetic order~\cite{Tran2018}, i.e., diverging at $T_\mathrm{N}$ and continuously decreasing at $T < T_\mathrm{N}$.
Such $\lambda_\mathrm{T}(T)$ suggests a very homogeneous distribution of local fields, consistent with the commensurate magnetic propagation vector in EuGa$_4$~\cite{Kawasaki2016}.
\tcr{At $T$ = 1.5\,K, in EuGa$_4$, $\lambda_\mathrm{T}$ is found 
to be $\sim$0.72\,$\mu$s$^{-1}$,} 
a value which is almost three orders of magnitude smaller 
than that of EuAl$_4$. This is also reflected in the ZF-$\mu$SR spectra shown in the insets of Fig.~\ref{fig:ZF_muSR}, where the damping of the muon-spin precession is much weaker in EuGa$_4$ than in EuAl$_4$.

The longitudinal $\mu$SR relaxation rates $\lambda_\mathrm{L}$ shown 
in Fig.~\ref{fig:ZF-EA}(c) and Fig.~\ref{fig:ZF-EG}(c) are much smaller
than the transverse relaxation rates $\lambda_\mathrm{T}$.
At 1.5\,K, $\lambda_\mathrm{L}$/$\lambda_\mathrm{T} \sim 0.15$ and 0.02 for EuAl$_4$ and EuGa$_4$, respectively.   
In contrast to $\lambda_\mathrm{T}(T)$ [see Fig.~\ref{fig:ZF-EA}(b) and Fig.~\ref{fig:ZF-EG}(b)], EuAl$_4$ and EuGa$_4$ exhibit a similar temperature-dependent $\lambda_\mathrm{L}(T)$, 
typical of materials with long-range magnetic order. 
In both cases, $\lambda_\mathrm{L}(T)$ 
diverges near $T_\mathrm{N}$, followed by a significant drop at $T < T_\mathrm{N}$, 
indicating that spin fluctuations are the strongest 
close to the onset of the AFM order. \tcr{At 1.5\,K,
$\lambda_\mathrm{L}$ is 8.2 and 0.01\,$\mu$s$^{-1}$ for EuAl$_4$ 
and EuGa$_4$, respectively}.
\tcr{In EuAl$_4$, at temperatures well inside the AFM state,} 
$\lambda_\mathrm{L}$ is hundreds of times 
larger than in EuGa$_4$, thus suggesting 
much stronger spin fluctuations in the AFM state of EuAl$_4$ 
than in EuGa$_4$. 
Conversely, in the PM state, both EuAl$_4$ and EuGa$_4$ exhibit 
similar $\lambda_\mathrm{L}$ values. 
Note that, in EuAl$_4$, as shown in Fig.~\ref{fig:ZF-EA}(d), muons implanted in the 
second site experience only the spin fluctuations. Consequently, 
$\lambda_\mathrm{tail}(T)$ in EuAl$_4$ shows similar features to 
$\lambda_\mathrm{L}(T)$, i.e., it exhibits a maximum near the onset of 
the AFM order and it, too, decreases as the temperature 
is lowered.
Future calculations of the muon-stopping sites, might 
be helpful to better appreciate 
the differences between EuAl$_4$ and EuGa$_4$. 
\tcr{In the fast-fluctuation limit (typical of magnetically ordered materials), 
the zero-field longitudinal muon-spin relaxation rate is described 
by:
\begin{equation}\label{eq:corela}
\lambda_\mathrm{L} = \frac{2\gamma_\mu^2\Delta^2}{\nu},
\end{equation} 
where $\Delta$ is the amplitude of field fluctuations, while $\nu$ is 
their correlation frequency (i.e., $1/\nu = \tau$, is the 
spin-correlation time)~\cite{Yaouanc2011}. 
The estimated spin-correlation times are $\tau$ = 1.3 and 8.2\,ns for EuAl$_4$ and EuGa$_4$, respectively. 
}

\begin{figure}[bht]
	\centering
	\vspace{-1mm}
	\includegraphics[width=0.48\textwidth]{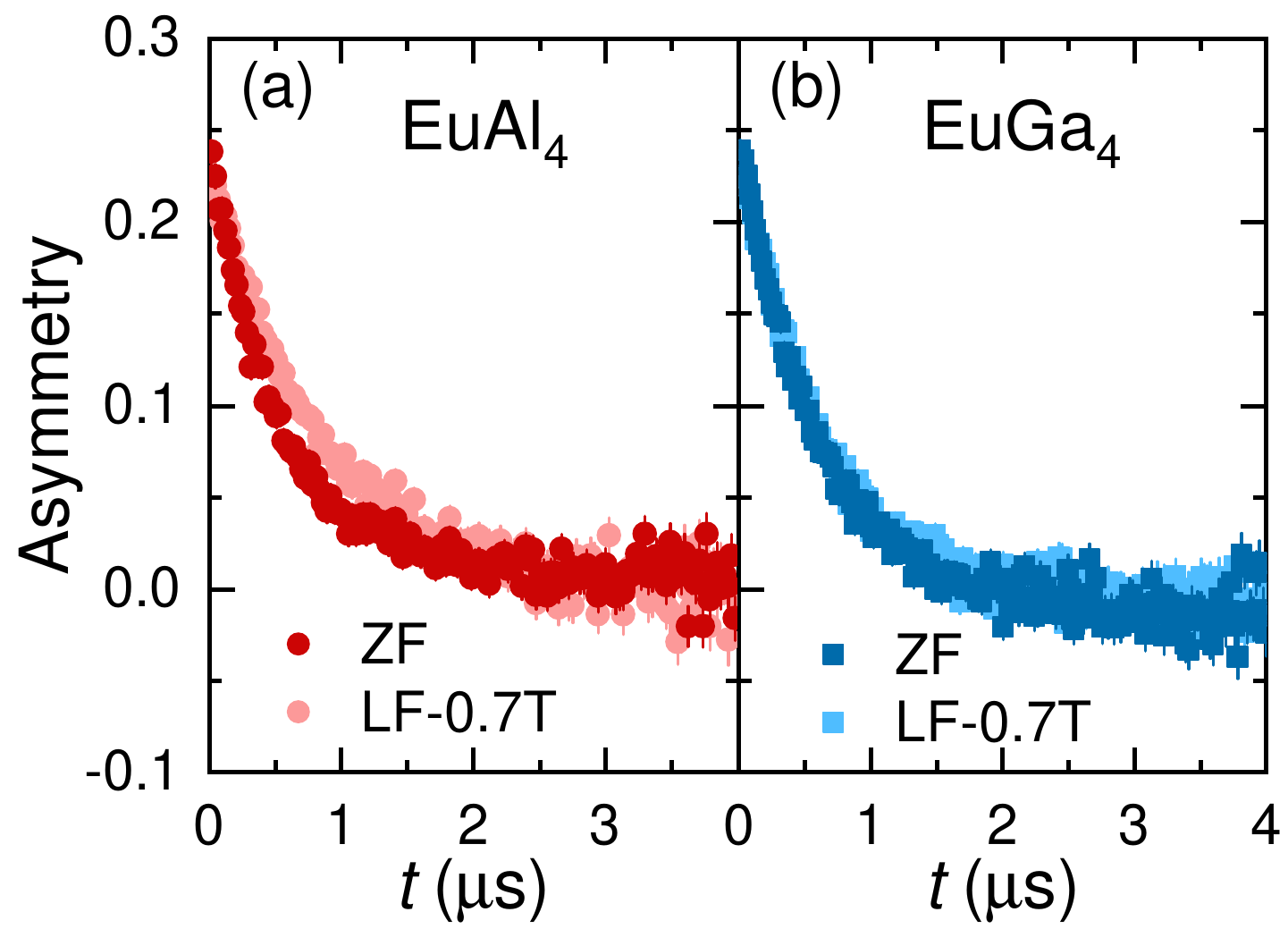}
	\caption{\label{fig:LF-muSR}LF-$\mu$SR time-domain spectra collected at 18\,K (i.e., slightly above $T_\mathrm{N}$) in an applied magnetic field of 0 and 0.7\,T in (a) EuAl$_4$ and (b) EuGa$_4$. Both spectra were collected in a longitudinal  muon-spin configuration, i.e., $\bm{p}_{\mu}$ $\parallel$ $\bm{S}_{\mu}$. The applied magnetic field is parallel to the muon-spin direction. 
	In either case, no appreciable 
	decoupling of muon spins with field can be identified. 	
}
\end{figure}

The vigorous 
spin fluctuations
in these compounds are
further supported by LF-$\mu$SR measurements. As shown in Fig.~\ref{fig:LF-muSR}, 
the $\mu$SR spectrum in a 0.7-T longitudinal field is almost 
identical to that collected in a zero-field condition, suggesting that 
muon spins cannot be decoupled, 
hence, that spin fluctuations 
survive even in a field of 0.7\,T in both EuAl$_4$ and EuGa$_4$.  
Note that, such spin fluctuations are robust against external 
magnetic fields, both in the AFM- (e.g., 1.5\,K) and in the PM state 
(i.e., 50\,K) far above $T_\mathrm{N}$ (see details in Fig.~\ref{fig:LF-muSR_EA} 
in the Appendix). 
Similar $\mu$SR results have been reported in other Eu-based materials, 
e.g., EuCd$_2$As$_2$, where the strong spin fluctuations cause 
the breaking of time-reversal symmetry and lead to the
formation of magnetic Weyl fermions~\cite{Ma2019}.  

\begin{table}[tbp]
	\renewcommand{\arraystretch}{1.2}
	\centering
	\caption{\label{tab:parameters} Summary of the EuAl$_4$ and EuGa$_4$ 
	single-crystal parameters obtained by means of magnetization- and $\mu$SR measurements.}
		\begin{tabular}{lcccccc} 
			\toprule
			                   & $T_\mathrm{N}^{\chi}$(K)  	& $T_\mathrm{N}^\mathrm{{\mu}SR}$(K)\footnotemark[1]    & $T_\mathrm{N}^\mathrm{{\mu}SR}$(K)\footnotemark[2]    &  $B_\mathrm{int}$(T)    & $\gamma$    &    $\delta$ \\
			\midrule 
			EuAl$_4^\mathrm{AFM1}$          & 15.6(2)                    & 13.9(1)                    & 16.0(2)                 & 0.57(5)    &   2.50(5)    &  0.50(5)  \\
		    EuAl$_4^\mathrm{AFM2}$          & 12.3(3)                    & ---                        & 13.7(4)                 & 0.33(2)    &   0.92(5)    &  0.17(3)  \\
		    EuGa$_4$                        & 16.5(2)                    & 15.2(3)                    & 16.5(4)                 & 0.89(2)    &   1.50(5)    &  0.50(5)  \\
			\bottomrule
		\end{tabular}	
		\footnotetext[1]{Determined from the asymmetry of wTF-$\mu$SR spectra.}
		\footnotetext[2]{Determined from fits of ZF-$\mu$SR spectra.}
\end{table} 

%
\subsection{Discussion}%
\tcr{
First we discuss why the successive magnetic transitions of EuAl$_4$ 
remain undetected by $\mu$SR, an absence which might be due to different reasons.
Firstly, the asymmetries obtained from 
wTF-$\mu$SR (see Figs.~\ref{fig:wTF} and \ref{fig:Vmag}) reflect the internal fields 
sensed by the implanted muons. However, when the applied
transverse field is much smaller 
than the internal fields, the wTF signal is mostly determined
by the muons implanted in the residual NM (or PM) fraction of a  
magnetically ordered sample. This is  reflected in 
a significant drop in the temperature-dependent asymmetry $A(T)$. 
In EuAl$_4$, below the onset of AFM order, the internal fields are 
hundreds of times larger than the applied wTF. Although changes in 
the magnetic structure, detected as successive transitions in the 
EuAl$_4$ magnetometry data, decrease the internal field from 
$\sim 0.4$\,T to 0.33\,T, this still remains much larger than wTF. 
Therefore, the successive magnetic transitions of EuAl$_4$ are not 
easily detectable via wTF-$\mu$SR measurements. 
Secondly, a slight change/rearrangement of the magnetic structure does not have a large impact on the internal field. According to neutron scattering studies, in EuAl$_4$, the magnetic $q$-vector changes from $\bm{q}_1$ = (0.085, 0.085, 0) at $T_\mathrm{N} = 13.5$\,K to $\bm{q}_2$  = (0.170, 0, 0) at 11.5\,K and slightly to $\bm{q}_3$  = (0.194, 0, 0) at 4.3\,K~\cite{Kaneko2021}. Therefore, we identify the drop of $B_\mathrm{int}$ at 13\,K with the critical temperature where the magnetic structure changes from $q_1$ to $q_2$. At the same time, the modification of magnetic structure from $q_2$ to $q_3$ with the magnetic moments pointing at the same direction is too tiny to have a measurable effect on $B_\mathrm{int}$.
Thirdly, changes in magnetic structure have little effect on the 
longitudinal relaxation rates $\lambda_\mathrm{L}$, which reflect 
solely the spin fluctuations in EuAl$_4$. In general, spin 
fluctuations decrease significantly as the temperature moves away 
from $T_\mathrm{N}$, but they diverge near the onset of the magnetic transition. 
Hence, in the magnetically ordered state, changes in 
$\lambda_\mathrm{L}$ caused by slight modifications of the magnetic structure are negligible compared to the temperature driven effects. 
}

\tcr{Since most of the skyrmion phases appear in a field range 
not easily accessible by standard $\mu$SR instruments, up to now, 
only a handful of results have been reported where LF-$\mu$SR is used 
to study the skyrmion compounds.
These include GaV$_4$(S,Se)$_8$~\cite{Franke2018}, Cu$_2$OSeO$_3$~\cite{Hicken2021}, 
and the Co-Zn-Mn alloy~\cite{Hicken2021,Ukleev2021}, whose skyrmion 
phases are stabilized by a relatively small field (< 0.1\,T). 
While for many  
newly discovered skyrmion systems, i.e., GdRu$_2$Si$_2$ and Gd$_3$Ru$_4$Al$_{12}$ (as well as for EuAl$_4$ and EuGa$_4$ studied here)~\cite{EuAl4_PRB,Shang2021,Khanh2020, Hirschberger2019}, the critical field required for stabilizing the skyrmion phase is above 1\,T.
In their AFM state, EuAl$_4$ and EuGa$_4$ exhibit comparable spin 
fluctuations to other well-studied skyrmion compounds. For instance, 
the muon-spin relaxation rates extracted from LF-$\mu$SR measurements in the skyrmion phases of 
Cu$_2$OSeO$_3$ and GaV$_4$(S,Se)$_8$ are $\sim 0.2$--0.8\,$\mu$s$^{-1}$, 
similar to those of Eu(Al,Ga)$_4$ (see Figs.~\ref{fig:ZF-EA} and \ref{fig:ZF-EG}). 
All these skyrmion compounds exhibit similar temperature-dependent 
muon-spin relaxation rates $\lambda_\mathrm{L}(T)$, with an enhanced and broadened 
peak in $\lambda_\mathrm{L}(T)$ at temperatures just below the critical temperature.  
Muon-spin relaxation rates also increase when entering the 
skyrmion phase by applying longitudinal magnetic fields, 
thus providing another method for identifying the presence of magnetic skyrmions. 
In the EuAl$_4$ and EuGa$_4$ case, where there is no skyrmion phase in 
zero field, the relaxation rates diverge at $T_\mathrm{N}$, 
followed by a significant drop at $T < T_\mathrm{N}$ due to the slowing 
down of spin fluctuations, a typical feature of magnetically ordered materials. 
A similar behavior is observed in Co$_{10}$Zn$_{10}$~\cite{Hicken2021}, 
a parent compound of the Co-Mn-Zn alloys, which lacks any skyrmion phases. 
According to Hall-resistivity measurements, the skyrmion phase may 
exist in a field range $\sim 1$--2.5\,T in EuAl$_4$ and 
$\sim 4$--7\,T in EuGa$_4$~\cite{EuAl4_PRB,Shang2021}. 
Aimed at investigating the intrinsic magnetic properties of both 
compounds, most of the current $\mu$SR studies are performed 
in zero-field conditions. 
To compare the muon-spin relaxation rates of EuAl$_4$ and EuGa$_4$ 
with those of other skyrmion compounds, and check if there are 
any skyrmion phases, further temperature-dependent $\mu$SR measurements 
under high magnetic fields are required. 
}


 %

The observation of a topological Hall effect in the magnetic state is
usually attributed to noncoplanar spin textures, such as magnetic
skyrmions, characterized by a finite scalar spin chirality in real space.
These magnetic skyrmions are stabilized by the Dzya\-lo\-shin\-skii\--Moriya
interaction, often observed in noncentrosymmetric materials~\cite{muhlbauer_skyrmion_2009,yu_near_2011,yu_real-space_2010,seki_observation_2012,kezsmarki_ne-type_2015,tokunaga_new_2015,Seki2012}. 
Conversely, magnetic materials with a centrosymmetric crystal structure that still host magnetic skyrmions are rare. To date, only a few systems have been reported, including 
some gadolinium intermetallic compound~\cite{kurumaji_skyrmion_2019,Hirschberger2019,Khanh2020},
Fe$_3$Sn$_2$~\cite{li_large_2019}, and possibly, also EuCd$_2$As$_2$~\cite{Xu2021}. 
In centrosymmetric systems, skyrmions can be stabilized,
for instance,
by magnetic frustration (e.g., in Gd$_3$Ru$_{4}$Al$_{12}$, Gd$_2$PdSi$_3$, and Fe$_3$Sn$_2$), 
or by the competition between the magnetic interactions and magnetic 
anisotropies (e.g., in GdRu$_2$Si$_2$)~\cite{Batista2016,Hirschberger2019,kurumaji_skyrmion_2019,li_large_2019,Khanh2020}.
According to magnetization and nuclear magnetic resonance studies, the magnetic anisotropy is moderate in EuAl$_4$ and EuGa$_4$~\cite{EuAl4_PRB,Shang2021,Niki2015}.  
Since both EuAl$_4$ and EuGa$_4$ adopt the same crystal structure of GdRu$_2$Si$_2$, skyrmions might be stabilized by the same mechanism. 
In addition, a four-spin interaction, mediated by itinerant electrons, 
has also been proposed as an important ingredient for the formation of 
skyrmions in centrosymmetric materials~\cite{Heinze2011,Batista2016,Ozawa2017}.  
Very recently, the chiral magnet Co$_7$Zn$_7$Mn$_6$ was found to host a 
skyrmion phase far below the magnetic ordering temperature, where  
spin fluctuations are believed to be the key for stabilizing the magnetic skyrmions~\cite{Ukleev2021}. Our $\mu$SR results reveal that both EuAl$_4$ and EuGa$_4$ exhibit
robust spin fluctuations against external magnetic fields, 
which analogously might be crucial for understanding the origin of 
topological Hall effect and of possible skyrmions in both materials.  

\section{Conclusion}
In summary, we investigated the temperature evolution of the local 
magnetic properties of EuAl$_4$ and EuGa$_4$ by means of $\mu$SR 
spectroscopy. wTF-$\mu$SR measurements confirm that EuAl$_4$ and EuGa$_4$
undergo an AFM transition at $T_\mathrm{N}$ $\sim$ 16 and 16.5\,K,
which are consistent with the magnetization data. The magnetic volume fractions, as determined from 
wTF-$\mu$SR asymmetry, are 91\% and 95\% for EuAl$_4$ and EuGa$_4$, 
respectively, implying a good sample quality in both cases.  
By using ZF-$\mu$SR measurements, we could follow the temperature evolution 
of the local magnetic fields and of spin fluctuations. The estimated 
internal fields at zero temperature are 0.33 and 0.89\,T for EuAl$_4$ and EuGa$_4$, respectively.  
EuAl$_4$ exhibits a more disordered internal field distribution than EuGa$_4$, reflected in a large transverse muon-spin relaxation rate $\lambda_\mathrm{T}$ far below $T_\mathrm{N}$, 
most likely related to its complex magnetic structure. The vigorous 
spin fluctuations revealed by both ZF-$\mu$SR and LF-$\mu$SR might be 
crucial for understanding the origin of
topological Hall effect and of possible skyrmions in EuAl$_4$ and EuGa$_4$.
In future, it might be interesting to investigate the magnetic properties 
of EuAl$_4$ and EuGa$_4$ using the $\mu$SR technique under high magnetic 
fields, where the topological Hall effect appears.  

\begin{acknowledgments}
T.S.\ acknowledges support from the Natural Science Foundation of 
Shanghai (Grant Nos.\ 21ZR1420500 and 21JC1402300) and the Schweizerische 
Nationalfonds zur F\"{o}rderung der Wis\-sen\-schaft\-lichen For\-schung 
(SNF) (Grant Nos.\ 200021\_188706 and 206021\_139082). Y.X.\ acknowledges 
support from the Shanghai Pujiang Program (Grant No.\ 21PJ1403100). 
This work was also financially supported by the National Natural Science 
Foundation of China (NSFC) (Grant Nos.\ 12174103 and 11874150) and the 
Sino-Swiss Science and Technology Cooperation (Grant No.\ IZLCZ2-170075).
We thank G.\ Lamura for the assistance during some phases of the 
LF-$\mu$SR experiments.
\end{acknowledgments}

\appendix
\section{Longitudinal-field $\mu$SR in EuAl$_4$}
%
\begin{figure}[bht]
	\centering
	\includegraphics[width=0.48\textwidth]{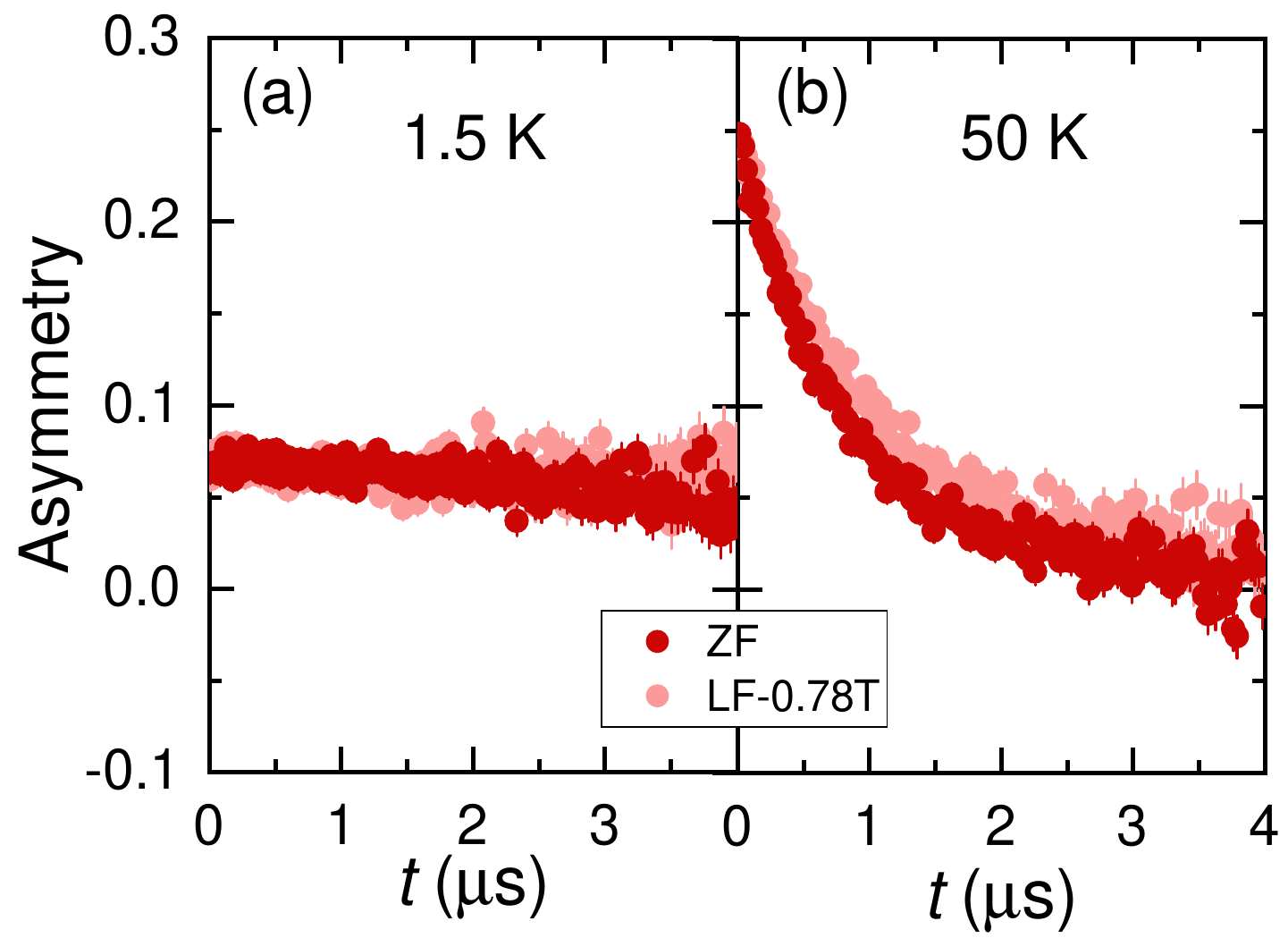}
	\caption{\label{fig:LF-muSR_EA}LF-$\mu$SR time-domain spectra collected at 1.5\,K (a) (far below $T_\mathrm{N}$) and 50\,K (b) (far above $T_\mathrm{N}$) in an applied magnetic field of 0 and 0.78\,T in EuAl$_4$. Both spectra were collected in a longitudinal muon-spin configuration, i.e., $\bm{p}_{\mu}$ $\parallel$ $\bm{S}_{\mu}$. The applied magnetic field is parallel to the muon-spin direction.  	
	}
\end{figure}
%
In Fig.~\ref{fig:LF-muSR_EA} we present the ZF- and LF-$\mu$SR spectra of EuAl$_4$, collected at temperatures well inside the AFM state 
(1.5\,K) and far above $T_\mathrm{N}$, in the PM state (i.e., 50\,K).
In the AFM state [see Fig.~\ref{fig:LF-muSR_EA}(a)], the fast drop of 
the $\mu$SR asymmetry reflects a very fast muon-spin depolarization 
in the first tenths of $\mu$s [see also ZF-$\mu$SR data in Fig.~\ref{fig:ZF_muSR}(a)].  
A 0.78-T longitudinal magnetic field has negligible effects 
on the long-time $\mu$SR spectra. Indeed, both the ZF- and LF-$\mu$SR 
spectra are almost identical, implying that the spin
fluctuations persist deep inside the AFM state of EuAl$_4$.
Surprisingly, similar features are observed also in the 
PM state, as clearly demonstrated in Fig.~\ref{fig:LF-muSR_EA}(b) [see also Fig.~\ref{fig:LF-muSR}].  
Since the data suggest that muon spins cannot be decoupled 
neither in the AFM nor in the PM state, this implies that, 
in this type of materials, \emph{spin fluctuations exist over a 
wide temperature range}, well above the AFM transition. 
We recall that, according to previous $\mu$SR studies on EuCd$_2$As$_2$, 
spin fluctuations are strongly enhanced
below 100\,K, thus causing the breaking of time-reversal symmetry 
and leading to the formation of magnetic Weyl fermions~\cite{Ma2019}.  
Further measurements at higher temperatures, including both ZF- and 
LF-$\mu$SR, are highly desirable to check if a similar
phenomenology occurs also in the BaAl$_{4}$-type family of materials.

\bibliography{EuA4.bib}

\end{document}